\documentclass[aps,prb,reprint,superscriptaddress]{revtex4-2}
\usepackage[colorlinks=true, linkcolor=blue, citecolor=blue,urlcolor=blue]{hyperref}
\newcommand{\scite}[1]{\textsuperscript{\cite{#1}}}

\usepackage{graphicx}
\usepackage{amsmath}
\usepackage{amssymb}
\usepackage{color}
\usepackage{float}
\newcommand{\SIQSE}{\affiliation{1}{Shenzhen Institute for Quantum Science and Engineering, Southern University of Science and Technology, Shenzhen, Guangdong, China}}
\newcommand{\DPHY}{\affiliation{2}{Department of Physics, Southern University of Science and Technology, Shenzhen, Guangdong, China}}
\newcommand{\IQA}{\affiliation{3}{International Quantum Academy, Shenzhen, Guangdong, China}}
\newcommand{\GDKL}{\affiliation{4}{Guangdong Provincial Key Laboratory of Quantum Science and Engineering, Southern University of Science and Technology, Shenzhen, Guangdong, China}}
\newcommand{\HFNL}{\affiliation{5}{
Shenzhen Branch, Hefei National Laboratory, Shenzhen 518048, China}}

\begin{document}
\title{M$^2$CS: A Microwave Measurement and Control System for Large-scale Superconducting Quantum Processors}
\author{Jiawei Zhang}
\thanks{These authors contributed equally to this work.}
\email{zhangjw2022@mail.sustech.edu.cn}
\affiliation{\SIQSE}\affiliation{\IQA}\affiliation{\GDKL}

\author{Xuandong Sun}
\thanks{These authors contributed equally to this work.}
\affiliation{\SIQSE}\affiliation{\IQA}\affiliation{\GDKL}\affiliation{\DPHY}

\author{Zechen Guo}
\affiliation{\SIQSE}\affiliation{\IQA}\affiliation{\GDKL}

\author{Yuefeng Yuan}
\affiliation{\IQA}
\author{Yubin Zhang}
\affiliation{\IQA}

\author{Ji Chu}
\affiliation{\IQA}

\author{Wenhui Huang}
\affiliation{\SIQSE}\affiliation{\IQA}\affiliation{\GDKL}

\author{Yongqi Liang}
\affiliation{\SIQSE}\affiliation{\IQA}\affiliation{\GDKL}

\author{Jiawei Qiu}
\affiliation{\SIQSE}\affiliation{\IQA}\affiliation{\GDKL}

\author{Daxiong Sun}
\affiliation{\SIQSE}\affiliation{\IQA}\affiliation{\GDKL}

\author{Ziyu Tao}
\affiliation{\IQA}

\author{Jiajian Zhang}
\affiliation{\SIQSE}\affiliation{\IQA}\affiliation{\GDKL}\affiliation{\DPHY}

\author{Weijie Guo}
\affiliation{\IQA}

\author{Ji Jiang}
\affiliation{\SIQSE}\affiliation{\IQA}\affiliation{\GDKL}

\author{Xiayu Linpeng}
\affiliation{\IQA}

\author{Yang Liu}
\affiliation{\IQA}

\author{Wenhui Ren}
\affiliation{\IQA}

\author{Jingjing Niu}
\affiliation{\IQA}\affiliation{\HFNL}

\author{Youpeng Zhong}
\email{zhongyp@sustech.edu.cn}
\affiliation{\SIQSE}\affiliation{\IQA}\affiliation{\GDKL}\affiliation{\HFNL}

\author{Dapeng Yu}
\affiliation{\SIQSE}\affiliation{\IQA}\affiliation{\GDKL}\affiliation{\DPHY}\affiliation{\HFNL}


\date{\today}

\begin{abstract}
As superconducting quantum computing continues to advance at an unprecedented pace, there is a compelling demand for the innovation of specialized electronic instruments that act as crucial conduits between quantum processors and host computers.
Here, we introduce a Microwave Measurement and Control System (M$^2$CS) dedicated for large-scale superconducting quantum processors. 
M$^2$CS features a compact modular design that balances overall performance, scalability and flexibility.
Electronic tests of M$^2$CS show  key metrics comparable to commercial instruments.
Benchmark tests on transmon superconducting qubits
further show qubit coherence and gate fidelities comparable to state-of-the-art results,
confirming M$^2$CS's capability to meet the stringent requirements of quantum experiments run on intermediate-scale quantum processors.
The system's compact and scalable design offers significant room for further enhancements that could accommodate the measurement and control requirements of over 1000 qubits, and can also be adopted to other quantum computing platforms such as trapped ions and silicon quantum dots.
The M$^2$CS architecture may also be applied to wider range of scenarios, such as microwave kinetic inductance detectors, as well as phased array radar systems. 
\end{abstract}
\maketitle
\section{Introduction}
Quantum computing holds the potential of solving some hard problems that are otherwise intractable with classical computers, such as large number factorization~\scite{Shor1,Grover1}. Superconducting qubits, one of the most promising platforms towards large-scale, fault-tolerant quantum computers, have made tremendous progresses in recent years~\scite{arute2019quantum,wu_strong_2021}. Both the quality and quantity of qubits in superconducting quantum processors have been rapidly improved~\scite{wu_strong_2021,xu_digital_2023,quafu}, breaking through the 1000 qubits barrier on a single superconducting quantum chip recently~\scite{IBM1000}. Concurrently, to achieve logical qubits and, ultimately, fault-tolerant quantum computing, some progress has been made in qubit error correction~\scite{acharyaSuppressingQuantumErrors2023,guptaEncodingMagicState2024,niBeatingBreakevenPoint2023,sivakRealtimeQuantumError2023}.
Such rapid progress poses new challenges for electronic instruments used to control and measure superconducting quantum processors.

Electronic instruments are crucial for precise control of superconducting qubits and high-fidelity measurement of their states. Taking the widely used transmon superconducting qubits as an example~\scite{koch2007,barends2013coherent}, electronic instruments must be able to perform the following functions at a minimum~\scite{bardin2020,krantz2019}: (1) output high-precision intermediate-frequency (IF) signals for qubit Z bias and adjust couplers between qubits; (2) output raido-frequency (RF) microwave signals (typically within the C band frequency range of 4--8 GHz) finely tuned for qubit XY operation and state detection; (3) sample RF microwave signals returned from the quantum processor, demodulate the signals to discriminate qubit states.
Driven by increasing demand in this field, commercial products have been developed to fulfill these requirements, however, their closed-source hardware and digital logic hinder custom optimization, and their high cost limits affordability. Laboratory-based electronic instruments have also been developed, some of which utilize a combination of customized RF hardware and commercial high-performance Field Programmable Gate Array (FPGA) evaluation boards with their FPGA logic being open-sourced~\scite{dingExperimentalAdvancesQICK2023,stefanazziQICKQuantumInstrumentation2022,xuQubiCExtensibleOpenSource2023,xuQubiCOpenSourceFPGABased2021}. Others employ entirely customized FPGA-based hardware to enhance scalability further~\scite{guoControlReadoutSoftware2019,linScalableCustomizableArbitrary2019,sunScalableSelfAdaptiveSynchronous2020,yangFPGAbasedElectronicSystem2022,yang2021fpga,wang_hardware_2021-1}.

\begin{figure*}[ht]
    \centering
    \includegraphics[width=0.9\textwidth]{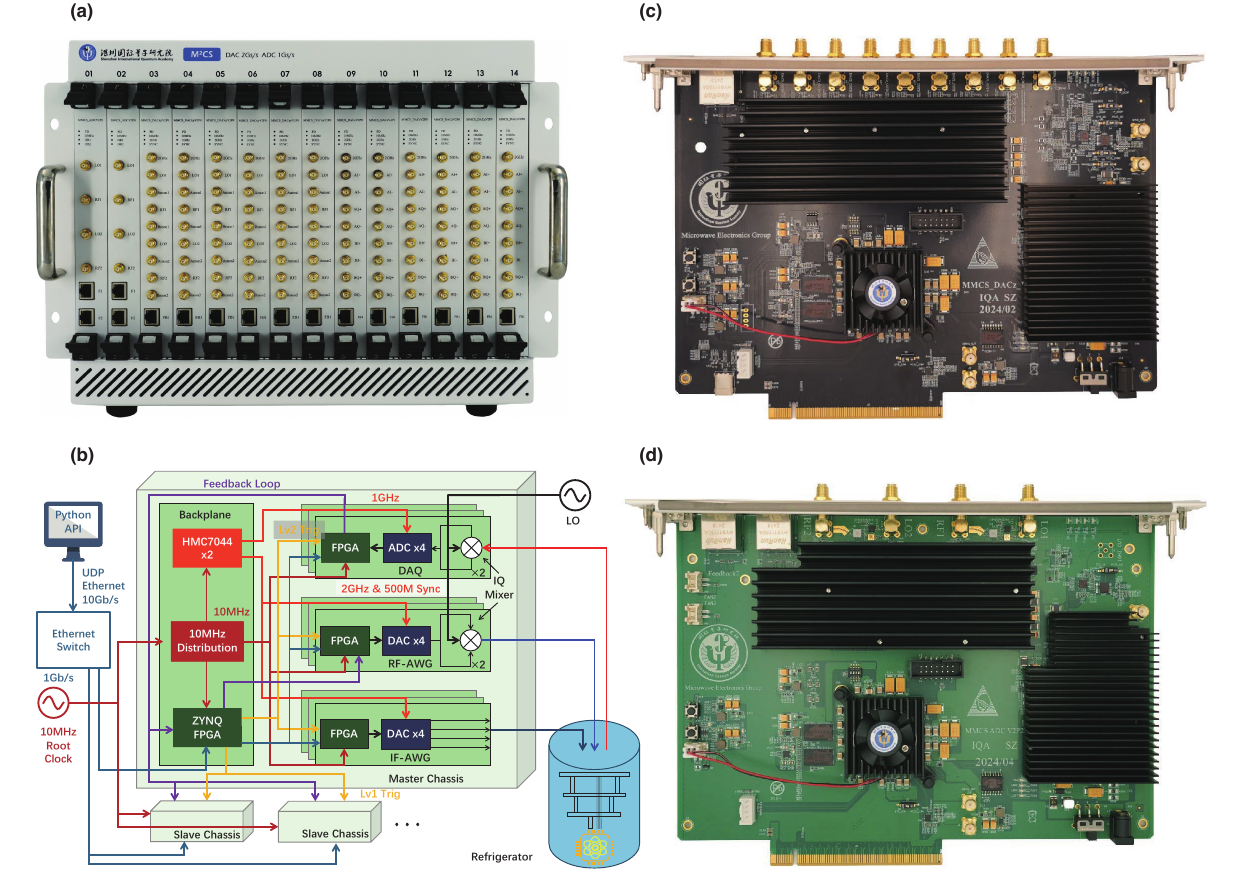}
    \caption{\textbf{\label{fig:architecture}System architecture of M$^2$CS.}
    (a) Photograph of a M$^2$CS chassis.
    (b) Schematic of the system architecture.
    (c) and (d) Photograph of an AWG and DAQ module respectively.
    }
\end{figure*}

Here, we introduce an FPGA-based electronic instrument dedicated for large-scale superconducting quantum processors--which we call the Microwave Measurement and Control System (M$^2$CS).
The M$^2$CS adopts a modular architecture for scalability and flexibility, where the hardware is housed within a 6U compactPCI chassis. Each chassis comprises one backplane and 14 slots which can accommodate arbitrary waveform generator (AWG) and data acquisition (DAQ) modules. Each AWG module features four digital-to-analog (DAC) channels that can generate IF pulses with a sampling rate of 2~Gsps and a vertical resolution of 14 bits. 
To generate RF microwave pulses for qubit XY control and state detection, some AWG modules are equipped with two surface-mounted In-phase and Quadrature (IQ) mixers that up-convert the IF pulses to RF pulses in the GHz range.
To avoid confusion, we call the AWG modules with or without IQ mixers RF-AWG or IF-AWG modules respectively, and refer to AWG when discussing features they share in common.
The DAQ modules are essentially the dual of RF-AWG modules.
Each DAQ module offers two RF input channels, which are down-converted to IF signals by the on-board IQ mixers, then digitized by analog-to-digital converters (ADCs) with a sampling rate of 1 Gsps and a resolution of 8 bits.
To meet expanded system requirements, multiple chassis can be cascaded in a tree-like structure, forming a scalable system with one master chassis connected to a maximum of 11 first-level slave chassis, and more than a hundred second-level slave chassis. 
The system's stability and efficiency are bolstered by a unified clock network, a 2-level trigger chain, and 1~Gbps UDP Ethernet communication, all integrated on the system backplane.
Electronics tests of M$^2$CS show key metrics comparable to commercial instruments.
The IF-AWG output has a spurious-free dynamic range (SFDR) of $<-50$~dBc over the entire bandwidth of 500~MHz, and a phase noise floor of $-140$~dBc/Hz beyond 10~kHz frequency offset. 
The IQ mixers on the RF-AWG modules can be carefully calibrated to suppress the carrier and image sideband leakage to below $-80$~dBm.
The DAQ module supports fast on-board demodulation of 12 channels with a feedback loop latency as low as 180~ns, enabling fast mid-circuit quantum measurement and error correction.
Finally, we benchmark the M$^2$CS performance with superconducting transmon qubits, showing $128.7$~$\mu$s qubit lifetime and 99.73\% CZ gate fidelity, comparable to state-of-the-art results.
This confirms M$^2$CS's capability to meet the stringent requirements of quantum experiments run on intermediate-scale superconducting quantum processors.

\section{System architecture}
\begin{figure*}[t]
    \centering
    \includegraphics[width=0.9\textwidth]{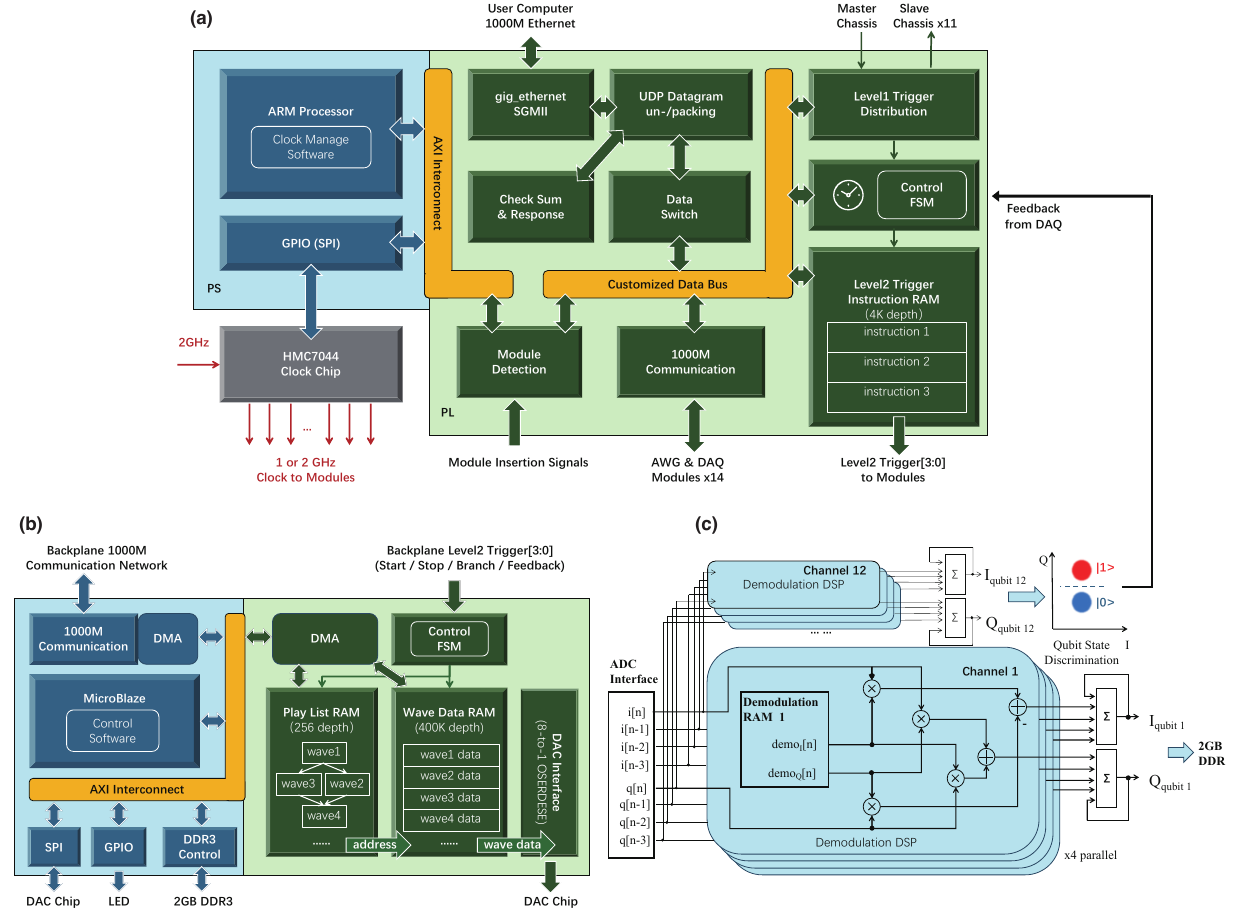}
    \caption{\label{fig:backplane_FPGA}
    \textbf{Block diagram of the backplane, the AWG and the DAQ modules.}
    (a) Schematic of backplane FPGA logic. The blue part is the Processing System (PS) side of ZYNQ, which includes an ARM processor. The processor is responsible for running clock management software and dynamically controlling the clock chip output at 1~GHz or 2~GHz depending on the DAQ or AWG module been detected. The green part is the Programmable Logic (PL) side of ZYNQ, which contains FPGA hardware resources. The FPGA is responsible for communication, triggering and feedback functions. The Level 1 trigger resets the timer inside the FSM, and when the timer value equals the timestamp in instructions, the Level 2 trigger is sent to the DAQ and AWG modules.
    (b) Schematic of AWG FPGA logic. The blue parts are responsible for data communication and register configuration before and after the experiment. The green parts are responsible for waveform generation during the experiment.
    (c) Schematic of DAQ DSP for real-time demodulation and qubit state discrimination.}
\end{figure*}


The system architecture of M$^2$CS, depicted in Fig.~\ref{fig:architecture}, comprises four principal parts: a unified clock network, a two-level trigger chain, a low-latency feedback loop and a 1Gbps UDP Ethernet communication network. These parts are primarily constituted by FPGA internal logic located on the backplane and the modules, supplemented by auxiliary components such as clock and communication chips.

The unified clock network distributes 10~MHz reference clock sourced from a Rubidium atomic clock. All FPGAs internally multiply the 10~MHz clock to 250~MHz, serving as the primary operating frequency.
The 2~GHz and 1~GHz sampling clocks required for the AWG and DAQ modules are distributed from the backplane using a clock generation device HMC7044 from Analog Devices Inc.~\scite{analog_devices_hmc7044}, ensuring precise clock alignment across modules. Additionally, the HMC7044 provides a unified 500~MHz synchronous clock to the digital-to-analog (DAC) chips on the AWG modules, facilitating synchronization of all output channels without manual calibration.

To flexibly control the synchronization of signal generation and acquisition in M$^2$CS, we adopt a two-level trigger chain architecture. The level 1 trigger is used to align the zero-time points of all chassis. When a user command is received from the host computer, a software start trigger is sent to the master chassis, which subsequently distributes level 1 trigger signals to all slave chassis. This action initiates simultaneous operation of all backplanes within the system. Subsequently, the backplanes transmit level 2 triggers to different modules based on user-defined timestamps, allowing users to control the start time of each module with flexibility.

In the low-latency feedback loop, the FPGA on the DAQ module discriminates the qubit states based on the phase of the demodulated signals. These state discrimination results are transmitted back to the backplane encoded in Low-Voltage Differential Signaling (LVDS) format via physically independent Category 8 Ethernet cables. The FPGA on the backplane, based on these results, sends level 2 branch triggers to the AWG modules at the user-defined timestamps. The FPGA on the AWG module selects the corresponding branch waveform data for output based on the branch triggers and a playlist uploaded in advance.

During the communication with host computers, M$^2$CS is tasked with uploading waveform data to AWG modules and downloading sampled data along with demodulation outcomes from DAQ modules. The total data volume can reach magnitudes of MB or even GB per experiment cycle. Given the impact of communication bandwidth on experiment efficiency, the chassis are connected to the host computer through a commercial ethernet switch operating at 10Gb/s, where data is distributed with a 1Gb/s UDP Ethernet protocol within each chassis  supported by the ZYNQ FPGA~\scite{xilinx_zynq-7000_2018} on the backplane, where UDP datagrams are unpacked and distributed either to the designated module or the backplane itself, based on the datagram's content.


Diverging from standard compactPCI chassis backplanes, the M$^2$CS backplane not only provides standard functionalities such as power supply and wiring but also offers highly integrated features, including dynamic clock networks, communication switching, and custom triggers, as depicted in Fig.~\ref{fig:backplane_FPGA}(a). The main control FPGA utilizes Xilinx's ZYNQ-7045 chip~\scite{xilinx_zynq-7000_2018}. On its Processing System (PS) side, it accommodates Dual-core ARM Cortex-A9 processors running clock management software. Upon detecting the insertion of AWG or DAQ modules, the ZYNQ controls the HMC7044 clock chip~\scite{analog_devices_hmc7044} to provide 2~GHz or 1~GHz clock for the module respectively. On the Programmable Logic (PL) side, hardware logic circuits are employed to perform unpacking, packing, and forwarding of 1Gbps UDP datagram. The low-latency nature of hardware logic ensures forwarding delays of only a few tens of nanoseconds.

In the level 2 trigger RAM, users can store custom 36-bit trigger instructions. This instruction set consists of lower 4 bits representing trigger types (start, stop, or branch requests for AWG modules to select branch waveforms) and upper 32 bits denoting timestamps. When the system timer matches the timestamp value, the control Finite State Machine (FSM) executes the instruction. For start/stop/branch trigger types, the FSM directly transmits the lower 4 bits of the instruction. For feedback trigger types, the FSM adjusts the output to branch 0 or 1 trigger based on feedback from the DAQ.



The AWG modules in our system are equipped with AD9739 DAC chips with a sampling rate of 2~Gsps and a resolution of 14~bits~\scite{analogdevices_ad9739}. Xilinx's Kintex-7 FPGA ~\scite{xilinx_7_2020} is utilized as the main control FPGA, allowing each channel to store waveform data up to 400K, with a maximum 200~$\mu$s waveform for each channel. We employ the LVDS communication format between the AD9739 and FPGA to minimize transmission delays. 

As illustrated in Fig.~\ref{fig:backplane_FPGA}(b), we utilize Xilinx's MicroBlaze soft-core processor within the FPGA, executing control software written in C language. This software manages pre- and post-experiment data processing as well as parameter configurations. Upon receiving waveform data from the backplane, it is temporarily stored in the 2~GB DDR memory via the AXI bus before being efficiently transferred to each channel's RAM using the MicroBlaze-controlled DMA (direct memory access) module.

During experiments, main control of the AWG is transferred to the control FSM. Trigger signals received from the backplane prompt the FSM to access the Play List RAM, retrieving the storage address of the next waveform data. Subsequently, waveform data is fetched from the Wave Data RAM based on this address and then promptly dispatched to the AD9739 via the interface module. To support branching functions, the Play List RAM accommodates waveform data storage structures with branches. This readout link has been finely optimized, with an internal FPGA delay of only 32~ns (8 clock cycles) from receiving the trigger to transmitting the waveform.

Each DAQ module features two dual-channel, 1~Gsps, 8-bit ADC08D1020 chips~\scite{texasinstruments_adc08d1020_nodate}. It also uses the Xilinx Kintex-7 FPGA as its primary control FPGA.
Each dual-channel ADC chip connects to an IQ mixer to down-convert the RF input to IF signals for subsequent sampling and processing. 
For low-latency demodulation of qubit readout signal, a digital signal processing (DSP) module has been designed within the DAQ FPGA, as depicted in Fig.~\ref{fig:backplane_FPGA}(c). Each RF input is down-converted to two IF signals first, which are subsequently digitized to two sequences of data $i[n]$ and $q[n]$ respectively, where $n$ is the index of the data. To obtain the qubit-state-imparted phase information, the data array are demodulated into a data point $(I,Q)$ in the phase space by the DSP in the following way:
\begin{equation} \label{eq:1-3}
\left\{ {\begin{array}{*{20}{c}}
{I = \sum\limits_n {i[n] \times demo{d_I}[n] - q[n] \times demo{d_Q}[n], } }\\
{Q = \sum\limits_n {i[n] \times demo{d_Q}[n] + q[n] \times demo{d_I}[n], } }
\end{array}} \right.
\end{equation}
where $demod_I [n]$ and $demod_Q [n]$ are user-defined demodulation factors, typically taking the form of
\begin{equation} \label{eq:1-2}
\left\{ {\begin{array}{*{20}{c}}
{demo{d_I}[n] = \cos ( - {\omega _{demod}}\times n\Delta t+\phi_{demod})},\\
{demo{d_Q}[n] = \sin ( - {\omega _{demod}}\times n\Delta t+\phi_{demod}),}
\end{array}} \right.
\end{equation}
where $\omega _{demod}/2\pi$ is the demodulation frequency, $\Delta t=1$~ns is the time step determined by the DAQ sampling rate, and $\phi_{demod}$ is the demodulation phase.
One can also incorporate an envelope such as Hann window into the demodulation factor to optimize the demodulation performance.

\begin{figure*}[t]
    \centering
    \includegraphics[width=0.9\textwidth]{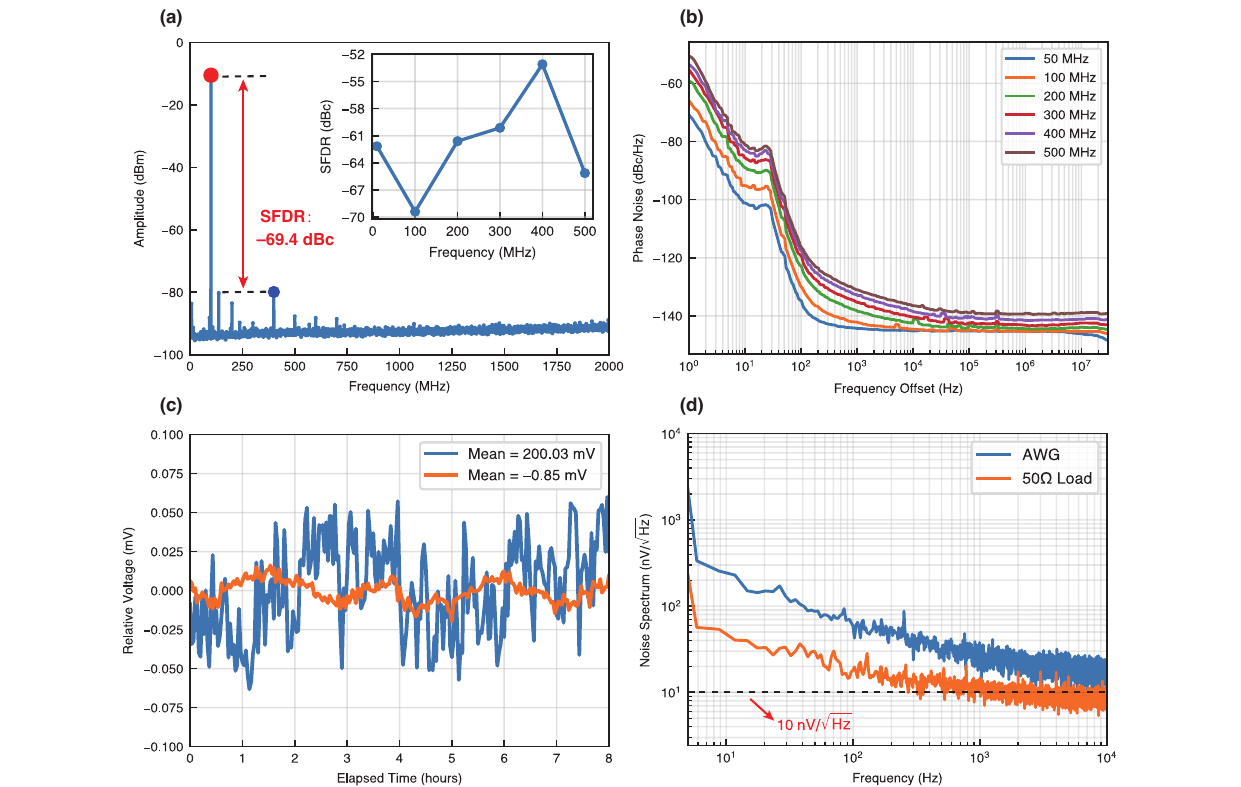}
    \caption{\textbf{IF-AWG module benchmark.} (a) AWG output spectrum, where a SFDR of $-69.4$~dBc is measured for a 100~MHz sinusoidal wave generated by the AWG. Inset: SFDR over the entire bandwidth of 500~MHz. (b) Phase noise measurement for different sinusoidal wave frequencies, reaching a phase noise floor of about $-140$~dBc/Hz at 10~kHz offset. (c) Voltage drift over 8 hours where the AWG output is set to 0~V (200~mV), showing an average voltage of $-0.85$~mV ($200.03$~mV) and a peak-to-peak fluctuation of 31~$\mu$V$_{\textrm{p-p}}$ (121~$\mu$V$_{\textrm{p-p}}$). Each data point is averaged over 1 minute. (d) Low frequency noise spectrum of the AWG output, reaching a noise floor of 20~$\textrm{nV}/\sqrt{\textrm{Hz}}$. Note the instrument noise floor is about 10~$\textrm{nV}/\sqrt{\textrm{Hz}}$, as confirmed by a 50~$\Omega$ load.}
    \label{fig:AWG}
\end{figure*}

Up to 12 channels of user-defined demodulation factors are supported for multiplexed readout of superconducting qubits coupled to  the same readout line.

The DAQ module also supports multiple-time readouts with up to 8~$\mu$s sample length. The module includes 2~GB of DDR memory, providing storage for 60,000 sets of demodulation results and over 10~ms of raw data storage per readout line.
To implement fast feedback control protocols, the DSP module compares the IQ demodulation data with the user-defined state discrimination thresholds to discriminate the superconducting qubit's state, which is subsequently transmitted to the backplane through the low-latency feedback loop.



\section{Electronics performance}


\begin{figure*}[t]
    \centering
    \includegraphics[width=0.9\textwidth]{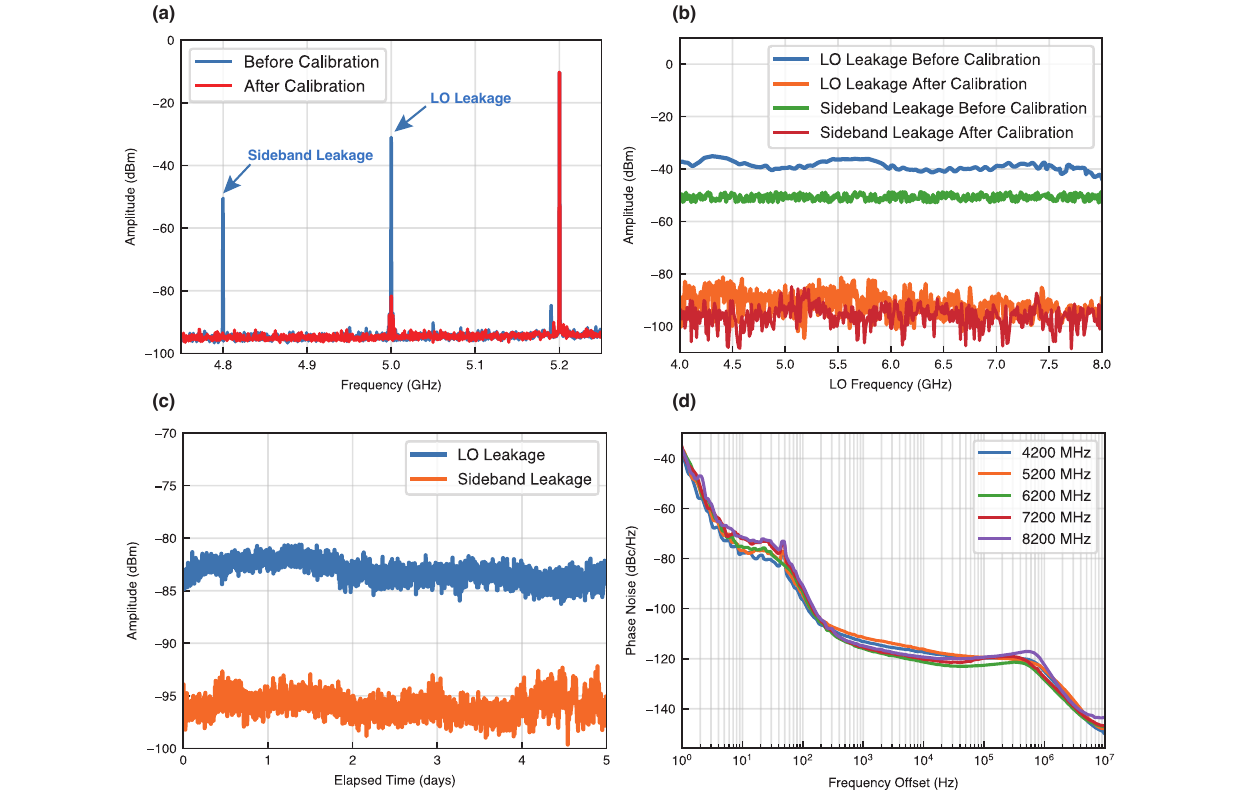}
        \caption{
        \label{fig:mixer}
        \textbf{RF-AWG module benchmark.}
        (a) RF-AWG output spectrum before and after calibration, where the LO frequency $\omega_{LO}/2\pi = 5$~GHz, and the sideband frequency $\omega_{sb}/2\pi = 200$~MHz. The LO leakage is suppressed from $\sim -30$~dBm to $<-80$~dBm, whereas the image sideband leakage is suppressed from $\sim -50$~dBm to $<-90$~dBm.
        (b) The LO and sideband leakage calibration at various $\omega_{LO}/2\pi$ from 4 to 8~GHz, with $\omega_{sb}/2\pi$ fixed at 200~MHz.
        (c) Long term stability of the mixer calibration.
        (d) Phase noise of the RF-AWG output, where$\omega_{LO}/2\pi$ are changed from 4 to 8~GHz at a step of 1~GHz, and $\omega_{sb}/2\pi$ is fixed at 200~MHz.
        }
\end{figure*}

\begin{figure*}[t]
    \centering
    \includegraphics[width=0.9\textwidth]{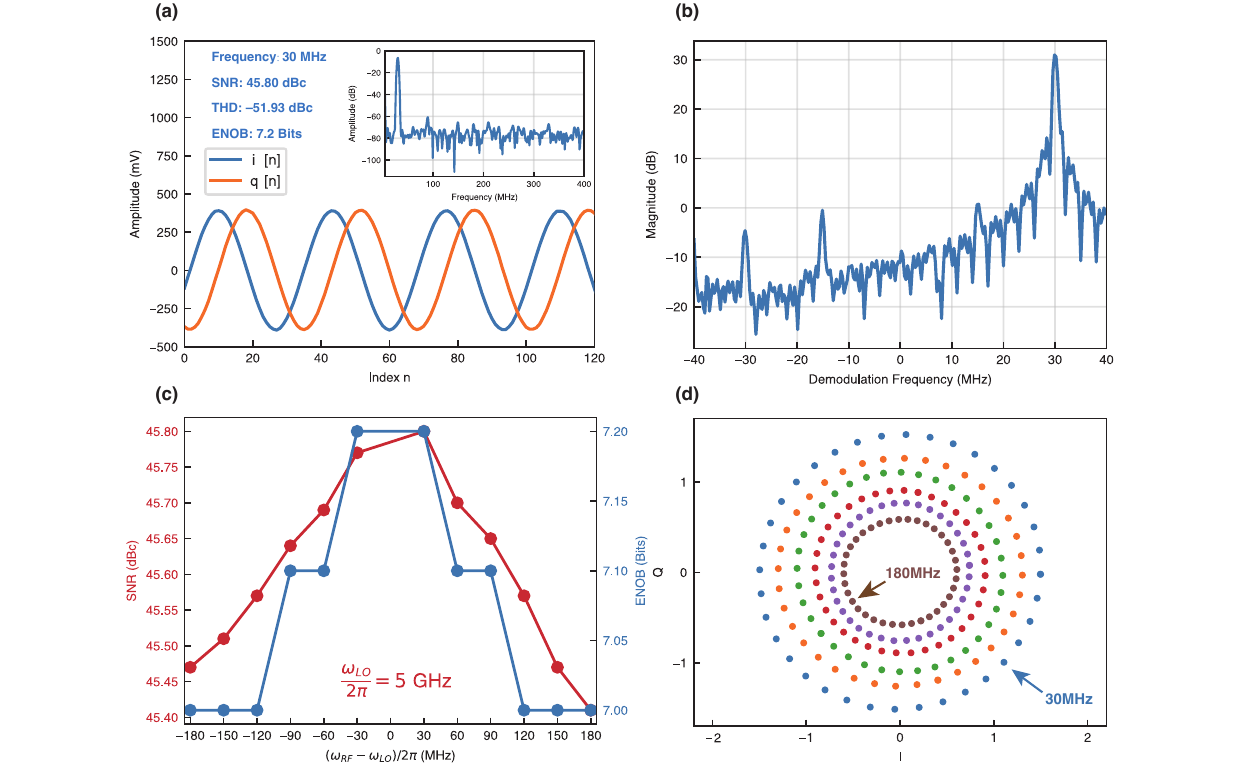}
    \caption{\textbf{DAQ module benchmark.} (a) Detecting an RF signal with frequency $\omega_{RF}/2\pi=5.03$~GHz. Setting the LO frequency of the DAQ to 5~GHz, the RF signal is down-converted to 30~MHz and then digitized, where the in-phase sequence $i[n]$ and quadrature sequence $q[n]$ differs by 90 degrees, as expected. Inset: FFT analysis of the $i[n]$ sequence, from which we obtain the SNR, THD and ENOB parameters. 
    (b) Fast demodulation of the sequence in (a) implemented on the FPGA. The demodulation frequency $\omega_{\rm demod}/2\pi$ is artificially varied here to demonstrate its frequency selectivity. A clear peak is observed in the demodulated signal magnitude when $\omega_ {\rm demod}/2\pi=30$~MHz.
    (c) SNR and ENOB of the DAQ for input RF signals of various frequencies $\omega_{RF}/2\pi$ from $(5-0.18)$~GHz to $(5+0.18)$~GHz, with $\omega_{LO}/2\pi$ set to 5~GHz. (d) Multiplexed demodulation. The input signal consists of a composite of various sideband frequency waves, ranging from 30 MHz to 180 MHz, with increments of 30 MHz. The input signal is digitized and simultaneously demodulated with $\omega_{\rm demod}/2\pi$ matching the sideband frequencies, while $\phi_{\rm demod}$  is varied at a step of 30 degrees. The demodulated data $(I,Q)$ for each $\omega_{\rm demod}/2\pi$  are concentric around the origin in the phase space.}
    \label{fig:DAQ}
\end{figure*}


\begin{figure*}[t]
    \centering
    \includegraphics[width=0.75\textwidth]{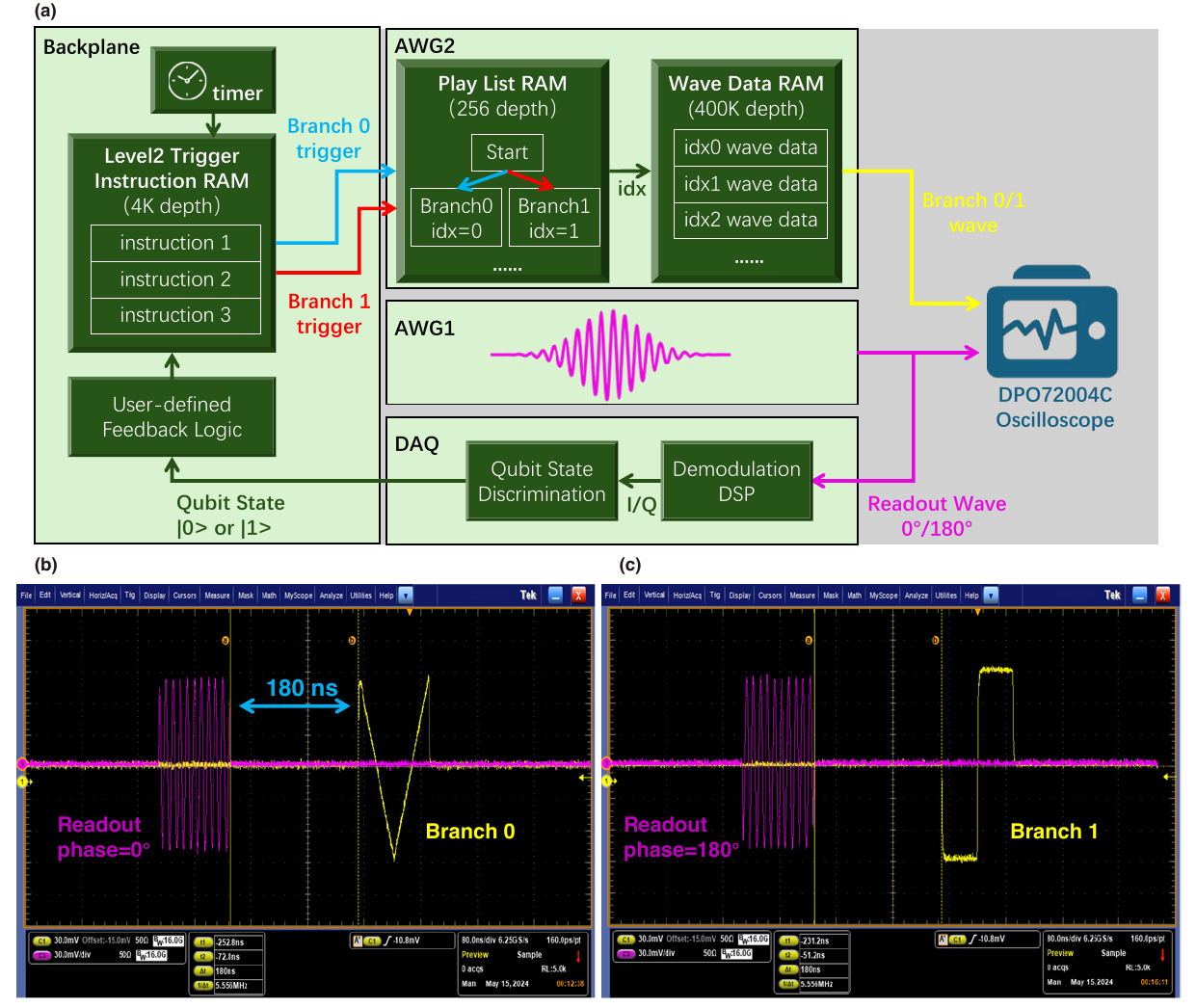}
    \caption{\textbf{Feedback loop test.} (a) Schematic of the feedback FPGA logic and the feedback loop test setup.
    (b) and (c) Oscilloscope result capturing that AWG2 outputs a saw or a square wave depending on the phase of the readout pulse generated by AWG1. A latency of 180~ns is observed from the end of the readout pulse generated by AWG1 to the start of the AWG2 output.}
    \label{fig:feedback}
\end{figure*}

The Spurious-Free Dynamic Range (SFDR) is a critical specification for evaluating the quality of IF-AWG output spectrum. SFDR quantifies the power distinction between the generated signal and the highest spurious signal, typically the harmonics.
We connect an IF-AWG module to a Rohde\&Schwarz FSL18 spectrum analyzer to measure its spectrum across the 10--2000~MHz range, see  Fig.~\ref{fig:AWG}(a) for example, where a SFDR of $-69.4$~dBc is measured for a 100~MHz sinusoidal wave generated by the IF-AWG.
The SFDR measurements at different IF-AWG output frequencies are shown as inset to Fig.~\ref{fig:AWG}(a), which are below $-50$~dBc over the entire bandwidth of 500~MHz. A Rohde\&chwarz FSWP26 phase noise analyzer was used to conduct the phase noise tests. We benchmarked the phase noise performance of sinusoidal waves generated by the IF-AWG at different frequencies from 50~MHz to 500~MHz, as depicted in Fig.~\ref{fig:AWG}(b). The results indicate a phase noise floor of about $-140$~dBc/Hz beyond 10~kHz frequency offset.
Integrating the phase noise in Fig.~\ref{fig:AWG}(b) from 1~Hz to 30~MHz yields a root mean square (RMS) jitter of 1~ps for the IF-AWG output.

Frequency tunable superconducting qubits require precise direct current (DC) sources to adjust their operating frequencies, as well as fast Z pulses to dynamically tune the qubit frequencies for gate operations.
As the number of qubits rapidly increases, it is highly desirable to reduce the wiring channels and use IF-AWGs to provide both the DC bias as well as fast Z pulses.
This requires proper wiring inside the dilution fridge, and that the IF-AWG output voltage is sufficiently stable to avoid qubit frequency drift.
The IF-AWG outputs are buffered by low noise, large-bandwidth differential amplifiers to convert the DAC differential outputs to single-ended outputs with a DC voltage swing of $-1$ to 1~V, yielding a least significant bit (LSB) resolution of 122~$\mu$V.
The differential amplifiers are powered by two stages of low-dropout regulators to reject ripples from the power supply.
As a benchmark test,
we set the IF-AWG output to 0~V and 200~mV respectively, and monitor its long term stability (8 hours) using a KEITHLEY DMM6500 6½ digital multimeter, showing a peak-to-peak fluctuation of 31~$\mu$V$_{\textrm{p-p}}$ and 121~$\mu$V$_{\textrm{p-p}}$ respectively, see Fig.~\ref{fig:AWG}(c).
With a typical wiring attenuation of 30~dB and a mutual inductance coupling of 2~pH to the qubit junction loops, this voltage fluctuation corresponds to a flux shift of the order of 1~$\mu\Phi_0$ and the qubit frequency shift of the order of 1~MHz, a tolerable value given the time span of 8 hours.
While the slow drift of the DC voltage only affects the operating frequencies of the qubits, the low frequency noise at the audio band can affect the phase coherence of frequency tunable qubits.
We measure the low frequency noise spectrum of the IF-AWG with a Rohde\&Schwarz upv audio analyzer, as shown in Fig.~\ref{fig:AWG}(d), where the spectrum reaches a noise floor of 20~$\textrm{nV}/\sqrt{\textrm{Hz}}$ at 10~kHz.
It is worth mentioning that the noise floor of the audio analyzer is 10~$\textrm{nV}/\sqrt{\textrm{Hz}}$, as confirmed by a 50~$\Omega$ load.

To generate RF microwave pulses for qubit XY control and state detection, RF-AWG modules are quipped with two surface-mounted IQ mixers.
IQ mixers can up-convert IF pulses that are relatively easy to
handle with electronic devices to RF microwave pulses, or inversely down-convert RF signals to IF signals for subsequent processing.
They have been commonly used in point-to-point communication,  test and measurement applications and so on.
An IQ mixer consists of four ports: the LO port which is driven by a local oscillator with a continuous drive of $\cos(\omega_{LO}t)$, two IF ports for in-phase (I port) and quadrature (Q port) modulation of the LO drive, and the RF port whose output is ideally given by $RF(t) = I(t)\cos(\omega_{LO}t) + Q(t)\sin(\omega_{LO}t)$.
By modulating the I and Q port at a sideband frequency as $I(t) = \cos(\omega_{sb}t)$, $Q(t) = -\sin(\omega_{sb}t)$, one can obtain a frequency up-conversion at the RF port: $RF(t) = \cos(\omega_{sb}t)\cos(\omega_{LO}t) - \sin(\omega_{sb}t)\sin(\omega_{LO}t) = \cos[(\omega_{LO}+\omega_{sb})t]$.
However, as nonlinear analog devices, nonidealities such as offset and amplitude/phase imbalance in the I and Q ports lead to signal leakage at the LO frequency or the image sideband frequency (i.e., $\omega_{LO}-\omega_{sb}$), which could detrimentally affect the qubit coherence and quantum gate fidelity.
Techniques that eliminate LO and image sideband leakage have been well-established at room
temperature utilizing spectrum analyzers or additional instruments for output diagnosis~\scite{jolin2020calibration,herrmann2022frequency}.
In Fig.~\ref{fig:mixer}(a), we show the RF-AWG output spectrum before and after applying such calibration techniques, where the LO frequency is 5~GHz, and the sideband frequency is 200~MHz. The LO leakage at 5~GHz is suppressed from $\sim -30$~dBm to $<-80$~dBm, whereas the image sideband leakage is suppressed from $\sim -50$~dBm to $<-90$~dBm.
Figure~\ref{fig:mixer}(b) shows the LO and sideband leakage calibration at various LO frequencies from 4 to 8~GHz, with the sideband frequency fixed at 200~MHz.
Both the LO and image sideband leakage are suppressed $<-80$~dBm after calibration over the entire C band.
One critical question arises as how stable is the calibration. In Fig.~\ref{fig:mixer}(c), we test the long term stability of the calibration and find the LO and image sideband leakage quite stable over 5 days.
Utilizing high speed DAC chips, one can also synthesize RF microwave pulses directly without the need for mixer calibration, but these cutting-edge devices usually have worse phase noise (typically around $-90$~dBc at 1~kHz offset) and complex synchronization.
The frequency mixing scheme we adopt here is relatively simple in system architecture because of low digital clock frequency and shows better phase noise performance.
Figure~\ref{fig:mixer}(d) shows the phase noise of the RF-AWG output, where the LO frequencies are changed from 4 to 8~GHz at a step of 1~GHz, the sideband frequency is fixed at 200~MHz, with a phase noise $<-110$~dBc at 1~kHz offset. This is a 20~dB improvement in terms of phase noise compared to direct synthesis schemes.
Limited by efficiency and resource, such calibration method is cumbersome for large-scale quantum processors requiring hundreds or even thousands of channels.
To address this challenge, an \emph{in situ} calibration method of mixers with superconducting qubits has been demonstrated~\scite{Wu2024}.

%



To test the functionality of the DAQ module, we drive the LO port of the mixers on the DAQ module at 5~GHz, and directly apply a continuous microwave signal to the RF input port.
As adjacent readout resonators are typically separated in frequency by $\sim 30$~MHz or more~\scite{arute2019quantum}, we apply a 5.03~GHz microwave signal to the RF input and digitize the down-converted IF signals to two sequences of data $i[n]$ and $q[n]$, as shown in Fig.~\ref{fig:DAQ}(a).
Fast Fourier transform (FFT) analysis of the $i[n]$ sequence is shown in inset of Fig.~\ref{fig:DAQ}(a), from which we obtain a signal-to-noise ratio (SNR) of 45.80 dBc, a total harmonic distortion (THD) of $-51.93$ dBc, and 7.2 effective number of bits (ENOB).
The FPGA on the DAQ module supports fast on-board demodulation of up to 12 channels, where the demodulation factors of each channel are user-defined.
In Fig.~\ref{fig:DAQ}(b), we show the fast
on-board demodulation of the signal in Fig.~\ref{fig:DAQ}(a), where the horizontal axis is the user-defined demodulation
frequency $\omega_{demod}/2\pi$ artificially varied to demonstrate its frequency selectivity. A clear peak is observed in the demodulated
signal magnitude when $\omega_{demod}/2\pi=30$~MHz, with more than 30~dBc contrast over spurious peaks.
We further vary the RF input frequency $\omega_{RF}/2\pi$ from $(5-0.18)$~GHz to $(5+0.18)$~GHz, and analyze the SNR and ENOB of the DAQ, as shown in Fig.~\ref{fig:DAQ}(c), with $>45$~dBc SNR and  $>7$~ENOB over the entire frequency range.
To verify the frequency multiplexing and phase coherence, we apply input waveforms comprising the superposition of sinusoidal waves at 6 different sideband frequencies from 30~MHz to 180~MHz with decreasing amplitudes. The input signal is digitized and simultaneously demodulated with demodulation frequencies $\omega_{demod}/2\pi$ matching the sideband frequencies, while the demodulation phase $\phi_{demod}$ is varied at a step of 30 degrees, see Fig.~\ref{fig:DAQ}(d). The demodulated data for the same $\omega_{demod}/2\pi$ are concentric around the origin in the phase space.


With the AWG and DAQ fully tested separately, we can connect them together and test the feedback functionality, as shown in Fig.~\ref{fig:feedback}.
In this test, AWG1 is used to generate a 100~ns readout wave to the DAQ module with phases of either 0 or 180 degrees to respectively simulate the detection of qubit \(|0\rangle\) or \(|1\rangle\) state. The DAQ's sampling window is also set to 100~ns to align with the readout waveform. 
After the DAQ completes the discrimination of the qubit state, the results are transmitted to the backplane via a 1~m Ethernet cable.

\begin{figure*}[t]
    \centering
    \includegraphics[width=0.9\textwidth]{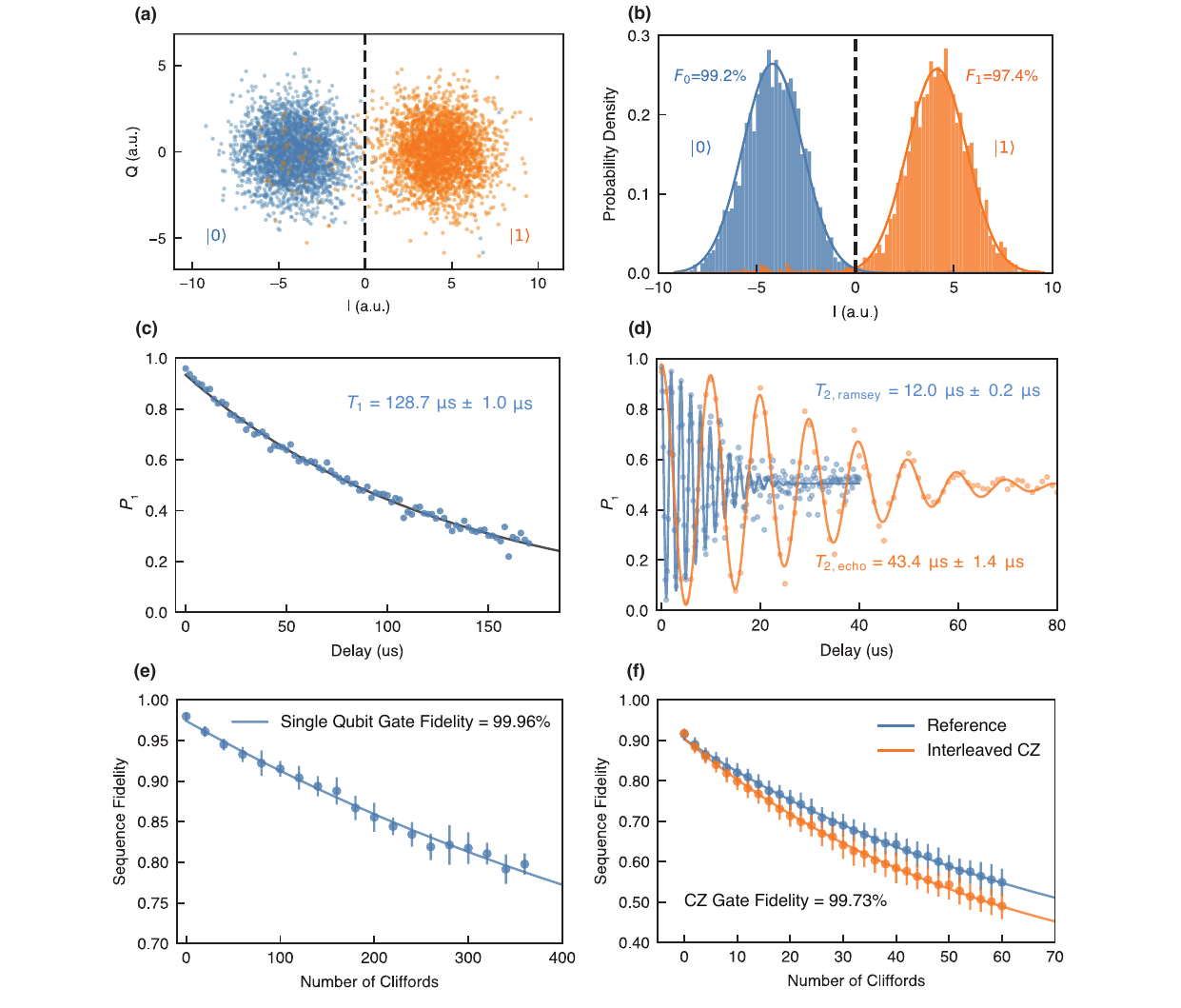}
    \caption{\label{fig:qubit} \textbf{Benchmarking M$^2$CS with superconducting qubits.}
    (a) IQ scatter plot, where the qubit is prepared in $|0\rangle$ and $|1\rangle$ state respectively. The dashed line is the state discrimination threshold. The in-phase (I) and quadrature (Q) data are expressed in arbitrary units (a.u.).
    (b) Histrogram of the data in (a) projected to the horizontal axis, with a readout fidelity of 99.2\% for the qubit ground state $|0\rangle$ and 97.4\% for the qubit excited state $|1\rangle$.
    (c), (d) Qubit coherence benchmark, showing a qubit \(T_1=128.7\)~$\mu$s, \(T_{2,ramsey}=12.0\)~$\mu$s and \(T_{2,echo}=43.4\)~$\mu$s.
    (e), (f) Qubit gate benchmark, showing a single qubit gate fidelity of 99.96\%, and two-qubit CZ gate fidelity of 99.73\%.
    }
\end{figure*}

The backplane's FPGA then dispatches branch trigger signals to AWG2 based on the outcomes of the state classification. AWG2 selectively outputs either branch 0 or branch 1 waveform accordingly.
AWG1 and AWG2 outputs are connected to the oscilloscope via equal-length cables. As illustrated in Fig.~\ref{fig:feedback}(b) and (c), the total closed-loop latency of the feedback is 180~ns (excluding the 100~ns sampling length). Specifically, the AWG latency is 72~ns, the backplane latency is 24~ns, the DAQ latency is 48~ns, and the remaining 36~ns comes from communication and wiring.

\section{Benchmark with superconducting qubits}

To demonstrate the practical performance of M$^2$CS, we utilize it to calibrate our superconducting quantum processor with 66 qubits~\scite{Yang2024}.
The readout resonators of the superconducting qubits are around 6~GHz.
As shown in Fig.~\ref{fig:qubit}(a) and (b), M$^2$CS achieves a readout fidelity of 99.2\% for the qubit ground state $|0\rangle$ and 97.4\% for the qubit excited state $|1\rangle$, without the assistance of parametric amplifiers.
We further benchmark the qubit coherence by measuring the lifetime \(T_1\), the Ramsey \(T_2\), and the echo \(T_{2,echo}\) of the qubits, as shown in Fig.~\ref{fig:qubit}(c) and (d). In the \(T_1\) measurement, an RF-AWG module outputs a so-called \(\pi\)-pulse at the qubit frequency after up-conversion, flipping the qubit from the ground state $|0\rangle$ to the first excited state $|1\rangle$. After some delay time \(t\), the qubit state is detected using the DAQ to discriminate the probability $P_1$ of the qubit being in the \(|1\rangle\) state. Scanning the delay time \(t\), we observe an exponential decay in $P_1$ with a decay constant of \(T_1=128.7\)~$\mu$s. In the \(T_2\) Ramsey measurement, a \(\pi/2\)-pulse from an RF-AWG rotates the qubit to an equal superpsition of \(|0\rangle\) state and \(|1\rangle\) state. After some delay time \(t\), another \(\pi/2\)-pulse is applied, followed by measurement using the DAQ. The probability $P_1$ of \(|1\rangle\) state exhibits an oscillation with evolution time \(t\), where the envelope of this oscillation decays at a dephasing time \(T_{2,ramsey}=12.0\)~$\mu$s. In the \(T_2\) spin echo measurement, an additional refocusing \(\pi\)-pulse is added between the two \(\pi/2\)-pulses to suppress low frequency noise, resulting in a longer coherence time of \(T_{2,echo}=43.4\)~$\mu$s.
We further calibrate single qubit gates and two-qubit control-Z (CZ) gates using randomized benchmarking (RB) to extract the gate errors by running $m$ cycles of random Clifford gates~\scite{arute2019quantum}, obtaining single qubit gate fidelity of 99.96\%, and two-qubit CZ gate fidelity of 99.73\%, see Fig.~\ref{fig:qubit}(e) and (f).
Both the qubit coherence and gate fidelities are comparable to state-of-the-art results.

\section{Conclusion}

In this study, we have demonstrated a Microwave Measurement and Control System (M$^2$CS) tailored specifically for large-scale superconducting quantum processors.
This customized and modular design can meet the stringent requirements of quantum experiments run on intermediate-scale quantum processors, effectively balancing overall performance, scalability and flexibility.
Electronic tests of M$^2$CS show comparable key metrics to those of commercial instruments.
Benchmark tests on transmon superconducting qubits confirm M$^2$CS's capability to control and measure quantum processors, where both the qubit coherence and gate fidelities are comparable to state-of-the-art results.
Moreover, the system's compact and scalable design offers significant room for further enhancements that could accommodate the measurement and control requirements of over 1000 qubits.
The M$^2$CS architecture may be adopted to other quantum computing platforms such as trapped ions~\scite{zhu2023fpga} and silicon quantum dots~\scite{xue2022quantum}. 
It could also be applied to a wider range of scenarios, such as Microwave Kinetic Inductance Detectors (MKIDs)~\scite{mchugh2012readout,stefanazziQICKQuantumInstrumentation2022}, as well as phased array radar systems.


\begin{acknowledgments}
{
 This work was supported by the Science, Technology and Innovation Commission of Shenzhen Municipality (KQTD20210811090049034, RCBS20231211090824040, RCBS20231211090815032), the National Natural Science Foundation of China (12174178, 12204228, 12374474 and 123b2071), the Innovation Program for Quantum Science and Technology (2021ZD0301703), the Shenzhen-Hong Kong Cooperation Zone for Technology and Innovation (HZQB-KCZYB-2020050), and Guangdong Basic and Applied Basic Research Foundation (2024A1515011714, 2022A1515110615).}
\end{acknowledgments}

\bibliography{MMCS}
\bibliographystyle{apsrev4-2}

\end{document}